\documentclass[a4paper, oneside, twocolumn, notitlepage, 10pt]{extarticle_ecoc}
\usepackage{ecoc}

\usepackage{xcolor}
\usepackage{float}
\AtEndPreamble{\DeclareSymbolFont{operators}{OT1}{\rmdefault}{m}{n}}
\usepackage[per-mode=symbol]{siunitx}
\usepackage{lipsum}
\usepackage{stfloats}

\begin{document}


\title{Real-time, Software-Defined, GPU-Based Receiver Field Trial}%


\author{Sjoerd~van~der~Heide\textsuperscript{(1,2,*)},
        Ruben~S.~Luis\textsuperscript{(1)},
        Benjamin~J.~Puttnam\textsuperscript{(1)},
        Georg~Rademacher\textsuperscript{(1)},\\
        Ton~Koonen\textsuperscript{(2)},
        Satoshi~Shinada\textsuperscript{(1)},
        Yoshinari~Awaji\textsuperscript{(1)},
        Chigo~Okonkwo\textsuperscript{(2)}, and
        Hideaki~Furukawa\textsuperscript{(1)}
}

\maketitle                  


\begin{strip}
 \begin{author_descr}
    
   \vspace{-2mm}\textsuperscript{(1)} NICT, 4-2-1 Nukui-Kitamachi, Koganei, Japan, \textsuperscript{(*)}\href{mailto:s.p.v.d.heide@tue.nl}{s.p.v.d.heide@tue.nl}

   \textsuperscript{(2)} High Capacity Optical Transmission Laboratory, Eindhoven University of Technology, the Netherlands
 \end{author_descr}
\end{strip}%
\setstretch{1.1}%
%
\begin{strip}
  \begin{ecoc_abstract}
    We demonstrate stable real-time operation of a software-defined, GPU-based receiver over a metropolitan network. Massive parallelization is exploited for implementing direct-detection and coherent Kramers-Kronig detection in real time at 2 and 1 GBaud, respectively.\vspace{-2mm}
  \end{ecoc_abstract}
\end{strip}


\section{Introduction}

\begin{figure*}[b!]
\vspace{-2mm}
\centering
\includegraphics{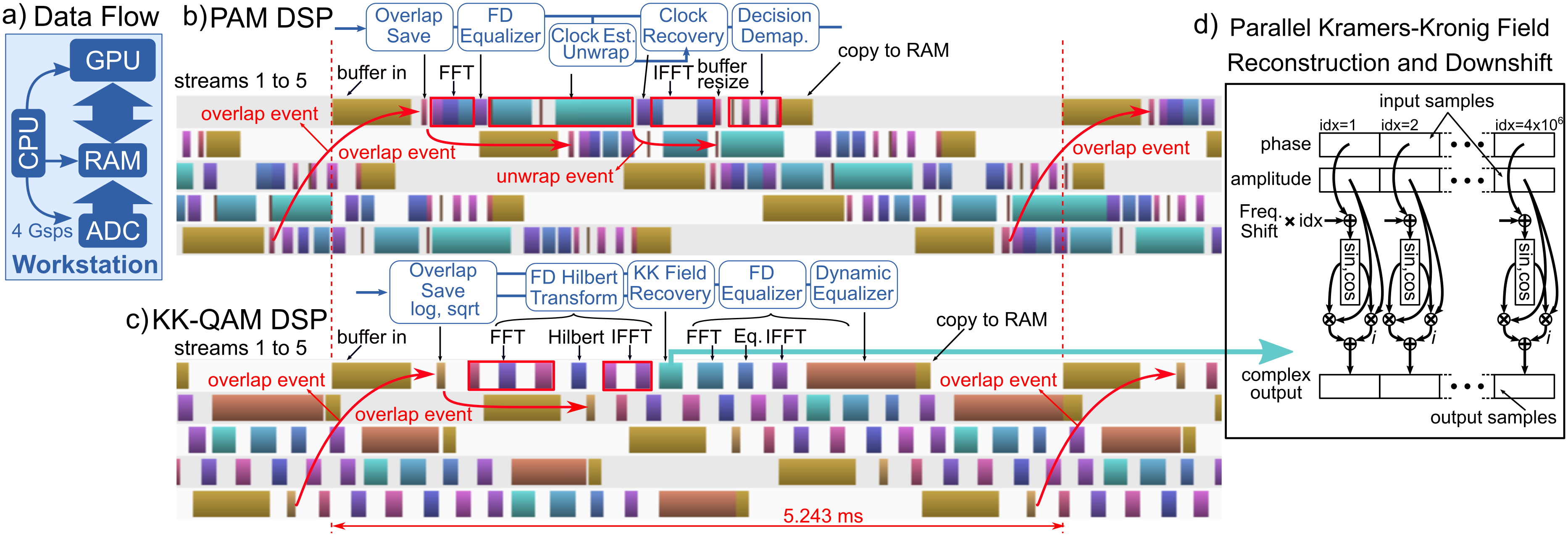}
\caption{a) Data flow within the workstation; DSP chains with annotated GPU profiler traces for (b) PAM and (c) KK-QAM signals. d) kernel for KK field reconstruction and frequency downshifting, showing the massive parallelism inside DSP blocks. Each stream processes buffers of 4,194,304 samples within 5.243~ms to keep up with the 4~Gs/s ADC, ensuring real-time operation.}
\vspace{-4mm}
\label{fig:dsp}
\end{figure*}


Software-defined transceivers are becoming commonplace for 5G and other wireless communications\cite{8949466}, avoiding costly electronics such as custom built ASICs. These systems perform digital signal processing (DSP) wholly\cite{GNURadio} or partially\cite{kazaz2017hardware} in off-the-shelf generic-purpose hardware, leading to high flexibility combined with low development effort and rapid turnaround. With ever increasing demand for optical data-traffic at lower cost-per-bit, interest in low-cost optical transceivers for inter-data-center networks has risen but energy and computing power limitations have thus far restricted development of software-defined transceivers for optical communications. However, with 45\% \cite{winzer_scaling_2017} year-on-year growth of computation power and 25\% \cite{sun_summarizing_2019} of energy efficiency (FLOPS per Watt), massive parallel processing of general purpose graphics processing units (GPUs) has the potential to meet this demand. Indeed, GPU power efficiency showed a 3-fold improvement over equivalent FPGAs for simple highly-parallelized operations \cite{qasaimeh_comparing_2019}, and, recently researchers have begun to explore the use of GPU-based DSP\cite{suzuki_software_2019,suzuki_real-time_2020}. However, there remains huge potential to exploit massive parallel processing on GPUs for low cost software-defined optical transceivers.

Here, we implement a software-defined multi-modulation format receiver with real-time DSP implemented on a commercial GPU and demonstrate transmission over a field deployed 91~km optical fiber link. The fiber ring is part of the JGN high speed R\&D network testbed\cite{JGN} consisting of 3 commercial reconfigurable add-drop multiplexers (ROADMs) in 2 separate Tokyo locations. The receiver DSP uses massive parallelization to receive 2-, 4- and 8-ary, pulse amplitude modulation (PAM) signals at 2~GBaud as well as quadrature phase-shift keyed (QPSK) and 16-ary quadrature-amplitude modulation (QAM) signals at 1~GBaud, with the latter detected using a Kramers-Kronig (KK) coherent receiver\cite{mecozzi_kramers_2016}. All measurements use identical transmitter and receiver hardware with format switching achieved by updating the GPU software. To the authors' knowledge, this is the first demonstration of a multi-modulation format software-defined GPU-based receiver and the first real-time demonstration of coherent KK detection. Further, we validate the performance over a field-deployed metropolitan network. These results show the potential of software-defined receivers for low-cost optical links, exploiting the exponentially growing computational power of GPUs.

\section{Multi-Format Real-Time DSP in GPU}

\begin{figure*}[b]
\centering
\includegraphics{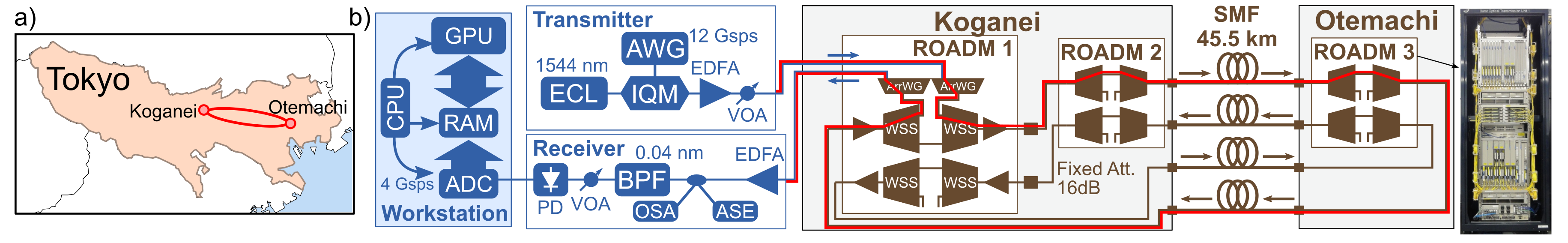}
\caption{Field-trial. a) Link map between Koganei and Otemachi; b) simplified experimental setup}
\label{fig:network}
\end{figure*}

Fig. \ref{fig:dsp}-a) shows the data flow within a workstation used for DSP. We used an ADC with a 1~GHz cut-off frequency, which took the form of a PCIe board. It sampled an incoming signal at 4~Gs/s, producing a continuous stream of buffers, each carrying 4,194,304 (2\textsuperscript{22}) 12-bit samples. These were sent to RAM and subsequently to the GPU memory for processing. A 3~GHz CPU controlled the data flow outside the GPU. After processing, the resulting binary signal was sent back to RAM for storage. Fig. \ref{fig:dsp}-b) and -c) show simplified diagrams of the DSP chains implemented in the GPU for processing of PAM signals and QAM signals using a Kramers-Kronig receiver. The latter are referred to here as KK-QAM. Unlike FPGA- or ASIC-based DSP, the required time for signal processing in GPUs is non-deterministic. As such, fluctuations between processing times of signal buffers within the GPU are handled by triggering events between parallel processing streams. In our system, each buffer was assigned a processing stream operating in parallel with other streams. Streams processing consecutive buffers communicated through event signals where dependencies between compute kernels dictated. Particularly for kernels computing overlap and save, which were crucial to maintain signal continuity. Each overlap and save kernel waited for the corresponding kernel of the previous stream to signal its completion with an overlap event so that part of the previous buffer could be prepended to its buffer. This is illustrated in Fig. \ref{fig:dsp}-b) and -c), which show annotated traces of the GPU profiler when processing PAM and KK-QAM signals, respectively, detailing the overlap event triggering 5 parallel processing streams. In the PAM DSP, we used 1024-point 100\% overlap and save, performed with a fast Fourier transform (FFT) operation of 8192 blocks in parallel. Afterwards, a 503-tap static equalizer was used. Blockwise clock-phase estimation and unwrapping was then performed with the latter also triggered by the stream processing the previous buffer, to ensure clock-phase continuity. Frequency domain (FD) clock recovery was performed, followed by symbol decision, demapping and copying the binary data to RAM. These stages are also noted in Fig. \ref{fig:dsp}-b).

A similar DSP chain was used for the KK-QAM signals. The KK front-end required nonlinear square root and logarithm operations on the received signal at 4 samples per symbol \cite{mecozzi_kramers_2016}, which were performed in the overlap and save kernel. A FD Hilbert transform was conducted using a 1024-point FFT and IFFT sequence to recover the phase of the optical signal field. This term was mixed with the square-root term and frequency downshifted. The recovered complex signal was filtered by a 203-tap static FD equalizer supported by a 1024-point FFT. A 512-point IFFT downsampled the signal to 2 samples per symbol and a 4-tap dynamic time-domain decision-directed least-mean squares equalizer performed clock-phase and symbol-phase recovery as well as symbol decision and demapping. Similarly to the PAM DSP, these operations were performed for 8192 data blocks per buffer in parallel. To illustrate the extent of the parallel processing performed in the GPU, an example of the kernel for field recovery in the KK-QAM DSP chain is included in Fig. \ref{fig:dsp}-d). Over 4 million threads in parallel fetched the amplitude and phase of a sample from GPU memory, applied a frequency shift, and wrote the complex result to GPU memory for further processing by other kernels.

In Fig. \ref{fig:dsp}-b), a gap can be observed between "copy to RAM" and "buffer in", implying the stream is idle and ready to process the next buffer. Since the five parallel streams each finish processing 2\textsuperscript{22} samples well within 5.243~ms ($\frac{5 \cdot 2^{22}}{4 \cdot 10^9}$) and are ready to receive a new buffer from the ADC, real-time operation at 4~Gs/s is ensured. 

\section{Experimental Demonstration}

Fig. \ref{fig:network}-b) shows a simplified diagram of the experimental setup, including the field trial network. At the transmitter, the lightwave from a 500~kHz linewidth external-cavity laser (ECL) centered at 1542.92~nm was modulated by a single-polarization IQ modulator (IQM). The latter was driven by a 2-channel arbitrary waveform generator (AWG) operating at 12~Gs/s. KK-QAM signals were produced by biasing the IQM to minimum output and the AWG set to produce a baseband 1~GBaud QPSK or 16-QAM signal with a 0.01 roll-off root-raised cosine (RRC) pulse shaping combined with a tone at a frequency of 0.547~GHz. PAM signals were produced by biasing one of the IQM arms to mid-point and driving it with 2~Gbaud, 2-, 4- and 8-PAM and 0.5 roll-off RRC. The signal power was set by an erbium-doped fiber amplifier (EDFA) followed by a variable optical attenuator (VOA). The transmission network consisted of a bidirectional ring with 3 commercial ROADMs. Two ROADMS were installed in the same location in Koganei, Tokyo. The link between these ROADMs was relatively short and its loss was set to 16~dB using fixed attenuators. Both ROADMs were connected to a commercial ROADM in Otemachi, Tokyo by a 45.5~km, 4-fiber link. The transmission loss, including optical distribution frames, was 16.5~dB. 56\% of the fiber was installed in underground ducts and the remainder on areal paths and in the surface along railway tracks. The red line in Fig. \ref{fig:network}-b) shows the signal path along the network, with a total transmission distance of 91~km. Each ROADM had two line sides, each consisting of wavelength selective switches (WSS) and optical amplifiers for add/drop and express connections. In addition, arrayed waveguide gratings (ArrWG) were used for add and drop. Fig. \ref{fig:network}-b) shows a photograph of one of the commercial ROADMs. The receiver consisted of an EDFA pre-amplifier followed by a 0.04~nm band-pass filter (BPF). In addition, a noise loading setup was included with an amplified spontaneous emission (ASE) source and an optical spectrum analyzer (OSA) through a 2$\times$2 coupler. A VOA was used to set the power at the input of a photodetector (PD) with a 3-dB cut-off frequency of 1~GHz. The PD output was directed to the ADC for processing as described in the previous section.

\begin{figure*}[t]
\vspace{-1mm}
\centering
\includegraphics{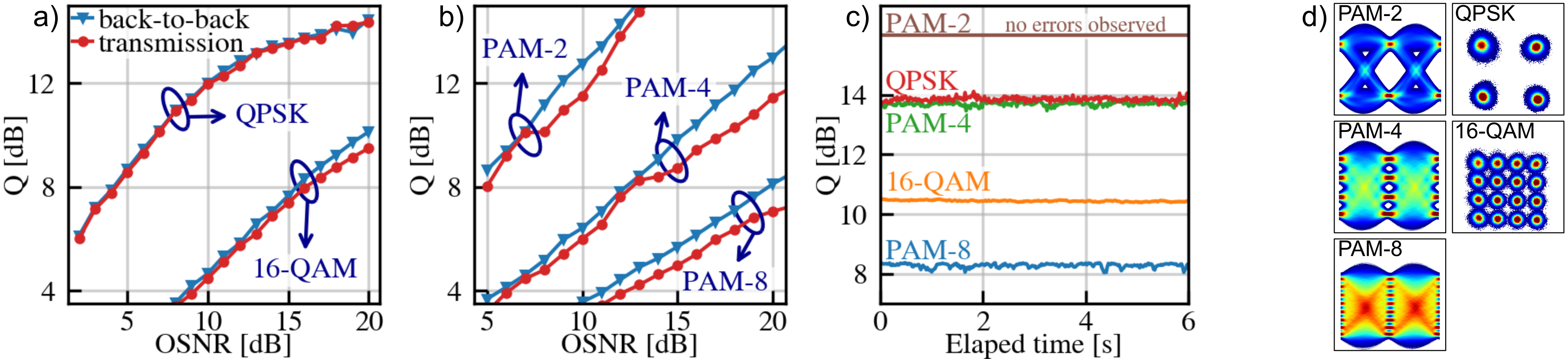}
\caption{Q-factor dependence on the OSNR in back-to-back and after transmission for KK-QAM (a) and PAM (b) formats. c) Q-factor time series after transmission during 6 seconds. d) Eye diagrams and constellations after transmission.}
\label{fig:results}
\end{figure*}

Fig. \ref{fig:results}-a) and -b) show the Q-factor dependence on the OSNR with 0.1~nm resolution bandwidth for KK-QAM and PAM signals, respectively, in back-to-back and after transmission. For KK-QAM signals, the OSNR includes the power of the tone for coherent KK detection. We used carrier-to-signal power ratios of 6~dB for QPSK and 11~dB for 16-QAM. This gave required OSNRs for Q = 8.4~dB of around 5~dB and 16~dB, respectively. The required OSNR for same Q was below 5~dB and around 14~dB for 2- and 4-PAM respectively. Hence, these formats are compatible with a hard-decision (HD) FEC with an overhead of 6.7\%\cite{agrell_information-theoretic_2018}. For 8-PAM, a Q-factor above 7~dB at an OSNR of 20~dB would allow a 20\% overhead HD-FEC\cite{agrell_information-theoretic_2018}. Fig. \ref{fig:results}-c) shows the Q-factor estimated in periods of 21~ms, for all considered formats after continuous transmission over a period of 6~seconds. During the measurement period, we observed no errors with 2-PAM. QPSK and 4-PAM showed similar performance with Q-factors around 14~dB. 16-QAM and 8-PAM reached Q-factor values around 10.5~dB and 8.5~dB, respectively. For all cases, we observed stable performance.

These results show the feasibility of software-defined receivers for optical communications exploiting parallel computation in GPUs and their interoperability with current commercial optical transmission systems for metropolitan reach. The continuous growth in processing capabilities and energy efficiency of GPUs heralds great potential for their future application in optical telecommunications systems.

\section{Conclusion}

We demonstrated stable real-time operation of a software-defined, GPU-based receiver over a field-deployed metropolitan network. We have shown the potential for massive parallel processing to recover in real time directly detected PAM signals as well as QAM signals with Kramers-Kronig coherent detection. The recovered signals were 2-, 4-, and 8-PAM at 2~GBaud and QPSK and 16-QAM at 1~GBaud. Performance was shown to remain stable despite the varying environment of installed fiber.
\\
\emph{\footnotesize \setstretch{1.1} Partial funding is from the Dutch NWO Gravitation Program on Research Center for Integrated Nanophotonics (Grant Number 024.002.033).}

\newpage
\printbibliography


\end{document}